\documentclass[aoas,MSNbibl,nameyear,rotating,dvips]{arximspdf}
\usepackage{dcolumn,multirow}
\usepackage{graphicx}

\doi{10.1214/12-AOAS580} 
\volume{7}
\issue{1}
\pubyear{2013}
\firstpage{177}
\lastpage{200}

\makeatletter
\newcolumntype{d}[1]{D{.}{.}{#1}}
\newcommand{\Db}{\mathbf{D}} 
\newcommand{\Yb}{\mathbf{Y}} 

\newcommand{\bb}{\mathbf{b}}
\newcommand{\wb}{\mathbf{w}}
\newcommand{\xb}{\mathbf{x}}
\newcommand{\yb}{\mathbf{y}}
\newcommand{\zb}{\mathbf{z}}

\newcommand{\alphab}{\bolds{\alpha}}
\newcommand{\betab}{\bolds{\beta}}
\newcommand{\epsb}{\bolds{\varepsilon}}
\newcommand{\mub}{\bolds{\mu}}
\newcommand{\phib}{\bolds{\phi}}
\newcommand{\psib}{\bolds{\psi}}
\newcommand{\thetab}{\bolds{\theta}}

\newcommand{\Sigmab}{\bolds{\Sigma}}

\newcommand{\Samp}{\mathcal{S}}

\newcommand{\logit}{\operatorname{logit}}

\newcommand{\MVN}{\mathcal{MVN}}

\renewcommand{\Pr}{\mathsf{P}}
\newcommand{\EV}{\mathsf{E}}
\newcommand{\var}{\mathsf{\operatorname{var}}}
\renewcommand{\vec}{\mathsf{\operatorname{vec}}}
\newcommand{\PED}{\mathsf{\operatorname{PED}}}
\newcommand{\iid}{\stackrel{\mathrm{i.i.d.}}{\sim}}

\newcommand{\tra}[1]{#1^\top}

\newcommand{\lbili}{\textsf{lbili}}
\newcommand{\platelet}{\textsf{platelet}}
\newcommand{\spiders}{\textsf{spiders}}
\newcommand{\PBCsub}{\textsf{PBC910}}
\makeatother

\begin{document}
\begin{frontmatter}

\title{Clustering for multivariate continuous and discrete longitudinal data}
\runtitle{Clustering for continuous and discrete longitudinal data}

\begin{aug}
\author[A]{\fnms{Arno\v{s}t} \snm{Kom\'arek}\corref{}\thanksref{T1}\ead[label=e1]{arnost.komarek@mff.cuni.cz}\ead[label=u1,url]{http://msekce.karlin.mff.cuni.cz/\textasciitilde komarek}}
\and
\author[B]{\fnms{Lenka} \snm{Kom\'arkov\'a}\ead[label=e2]{komarkol@fm.vse.cz}\thanksref{T2}}
\thankstext{T1}{Supported by the Czech Science Foundation Grant GA\v{C}R 201/09/P077.}
\thankstext{T2}{Supported by the Czech Science Foundation Grant GA\v{C}R P403/12/1557.}

\runauthor{A. Kom\'arek and L. Kom\'arkov\'a}
\affiliation{Charles University in Prague and University of
Economics in Prague}
\address[A]{Faculty of Mathematics and Physics\\
Charles University in Prague\\
Sokolovsk\'a 83 \\
CZ--186 75, Praha 8\\
Czech Republic\\
\printead{e1} \\
\printead{u1}}
\address[B]{Faculty of Management \\
University of Economics in Prague \\
Jaro\v{s}ovsk\'a 1117 \\
CZ--377 01, Jind\v{r}ich\r{u}v Hradec \\
Czech Republic \\
\printead{e2}}
\end{aug}

\pdfauthor{Arnost Komarek, Lenka Komarkova}

\received{\smonth{3} \syear{2011}}
\revised{\smonth{7} \syear{2012}}

%
\begin{abstract}
Multiple outcomes, both continuous and discrete, are routinely gathered
on subjects in longitudinal studies
and during routine clinical follow-up in general. To motivate our work,
we consider a longitudinal study
on patients with primary biliary cirrhosis (PBC) with a continuous
bilirubin level, a discrete platelet count
and a dichotomous indication of blood vessel malformations as examples
of such longitudinal outcomes.
An apparent requirement is to use all the outcome values to classify
the subjects into groups
(e.g., groups of subjects with a similar prognosis in a clinical setting).
In recent years, numerous approaches have been suggested for
classification based on longitudinal
(or otherwise correlated) outcomes, targeting not only traditional
areas like biostatistics, but also
rapidly evolving bioinformatics and many others. However, most
available approaches consider only continuous outcomes
as a basis for classification, or if noncontinuous outcomes are
considered, then not in combination
with other outcomes of a different nature. Here, we propose
a statistical method for clustering (classification)
of subjects into a prespecified number of groups with a priori unknown
characteristics on the basis of repeated measurements
of several longitudinal outcomes of a different nature.
This method relies on a multivariate extension of the classical
generalized linear mixed model
where a mixture distribution is additionally assumed for random effects.
We base the inference on a Bayesian specification
of the model and simulation-based Markov chain Monte Carlo methodology.
To apply the method in practice, we have prepared ready-to-use software
for use in R (\url{http://www.R-project.org}).
We also discuss evaluation of uncertainty in the classification and
also discuss
usage of a recently proposed methodology for model comparison---the
selection of a number of clusters in our case---based on the penalized
posterior deviance
proposed by Plummer [\textit{Biostatistics} \textbf{9} (2008) 523--539].
\end{abstract}

%
\begin{keyword}
\kwd{Classification}
\kwd{functional data}
\kwd{generalized linear mixed model}
\kwd{multivariate longitudinal data}
\kwd{repeated observations}.
\end{keyword}

\end{frontmatter}

\section{Introduction}\label{secIntro}

\subsection{Data and the research question}\label{subsecdataAndQuestion}
In clinical practice multiple markers of disease progression, both
continuous and discrete,
are routinely gathered during the follow-up to decide on future
treatment actions.
Our work is motivated by data from a Mayo Clinic trial on 312 patients
with primary
biliary cirrhosis (PBC) conducted between 1974--1984 [\citet
{Dicksonet89}]. This longitudinal study
had a median follow-up time of 6.3 years with a large number of
clinical, biochemical, serological
and histological parameters recorded for each patient.
The data are available in \citet{FlemingHarrington91}, Appendix~D, and
electronically at
\href{http://lib.stat.cmu.edu/datasets/pbcseq}{http://lib.stat.cmu.edu/datasets/}
\href{http://lib.stat.cmu.edu/datasets/pbcseq}{pbcseq}.

With these data, we shall mimic a common problem from the clinical practice:
at a prespecified time point from the start of follow-up we want to use
the values of the markers
of the disease progression to identify groups of patients with similar
characteristics. That is, we want to perform
a cluster analysis using the longitudinal measurements.
With these motivating data, we perform a classification of patients who
survived without liver
transplantation the first 910 days (2.5 years)
of the study ($N=260$), the data being further referred to as {\PBCsub}.
This corresponds to the practical problem outlined above, that is,
clustering of patients being available at a given time point.
For the purpose of this paper, the time point of 910 days was selected
arbitrarily.
Its choice in other application can, of course, be driven by practical
or other considerations.
The following markers will be considered for the cluster analysis:
continuous logarithmic serum bilirubin (\lbili),
discrete platelet count (\platelet) and dichotomous indication of blood
vessel malformations
(\spiders); see Figure~\ref{figPBC01}.

In a clinical routine usually only the last available measurements
reflecting the current patient status are used to identify
the prognostic groups---clusters.
Clearly, such a procedure ignores the available information on the
markers' evolution
over time, which might be more important for reasonable classification
than simply the last known state.
To remedy this deficiency, we shall propose a clustering method
exploiting jointly the whole history of longitudinal measurements
of all considered markers which might have a different nature from
being continuous to discrete, or even dichotomous.

\subsection{Basic notation and data characteristics}\label{subsecnotation}
Let $\Yb_{i,r} = (Y_{i,r,1},\ldots,\break Y_{i,r,n_{i,r}}\tra{)}$ denote
a random vector of the longitudinal profile
of the $r$th marker ($r=1,\ldots,R$) pertaining to the $i$th subject
($i=1,\ldots,N$).
Further, let $\Yb_i = (\tra{\Yb_{i,1}},\ldots,\tra{\Yb_{i,R}}\tra{)}$
be a random vector of all longitudinal measurements
on the $i$th subject and $\Yb= (\tra{\Yb_1},\ldots,\tra{\Yb_N}\tra{)}$
a random vector representing all available
outcomes. As usual, let $y_{i,r,j}, \yb_{i,r}, \yb_i, \yb$ denote the
observed counterparts
of corresponding upper case random variables and vectors.
Throughout the paper, we will assume the independence of subjects, that
is, independence of
$\Yb_1,\ldots,\Yb_N$.
Furthermore, let $n = \sum_{i=1}^N\sum_{r=1}^R n_{i,r}$ be the total
number of observations and
let $t_{i,r,j}$ be the times (on a study time scale) at which the
individual values $Y_{i,r,j}$ ($i=1,\ldots,N$, $r=1,\ldots,R$,
$j=1,\ldots,n_{i,r}$)
were taken. Finally, let $p(\cdot)$ and $p(\cdot| \cdot)$ be generic
symbols for (conditional) distributions.

In the {\PBCsub} data and our application, the number of markers $R$
equals~3. As it is common with the longitudinal data,
numbers $n_{i,r}$ of available measurements of each marker varies
(between 1 and 5) across patients (median~4), leading to $n = 2 734$.
Further, the distribution of the time points $t_{i,r,j}$ also varies
across subjects and markers; see Table~\ref{tab01}.
However, note that the fifth visit which is available for only three
patients is not outlying with respect to its timing from the rest of
the data set.
Indeed, it only corresponds to patients with a slightly more frequent
visiting schedule.
In summary, our longitudinal data are heavily unbalanced and
irregularly spaced in time, also containing 12 patients for whom
only baseline marker values at time $t = 0$ are available.
It is our aim to also classify or at least suggest a classification for
those patients.\vspace*{-2pt}

%
\begin{table}
\caption{Data {\PBCsub}. Characteristics of the time points at which
the longitudinal values of the markers for clustering were taken.
For each marker $r$ and each visit $j$, $n_{r,j}^*$ gives the number of
available measurements, med, $Q_1$ and $Q_3$ are the median,
the lower and the upper quartile of the time points in months when the
measurements were taken}\label{tab01}
\begin{tabular*}{\textwidth}{@{\extracolsep{\fill}}lccccccccc@{}}
\hline
& \multicolumn{3}{c}{{\textbf{\lbili}} $\bolds{(r=1)}$} & \multicolumn
{3}{c}{{\textbf{\platelet}} $\bolds{(r=2)}$} &
\multicolumn{3}{c@{}}{{\textbf{\spiders}} $\bolds{(r=3)}$}
\\[-4pt]
& \multicolumn{3}{c}{\hrulefill} & \multicolumn{3}{c}{\hrulefill} &
\multicolumn{3}{c@{}}{\hrulefill} \\
& & \multicolumn{2}{c}{$\bolds{t_{i,r,j}}$ \textbf{(months)}}
& & \multicolumn{2}{c}{$\bolds{t_{i,r,j}}$ \textbf{(months)}}
& & \multicolumn{2}{c@{}}{$\bolds{t_{i,r,j}}$ \textbf{(months)}} \\[-4pt]
& & \multicolumn{2}{c}{\hrulefill}
& & \multicolumn{2}{c}{\hrulefill}
& & \multicolumn{2}{c@{}}{\hrulefill} \\
$\bolds{j}$ & $\bolds{n_{r,j}^*}$ & \textbf{med} & $\bolds
{(Q_1$--$Q_3)}$ & $\bolds{n_{r,j}^*}$ & \textbf{med} &
$\bolds{(Q_1$--$Q_3)}$ & $\bolds{n_{r,j}^*}$ & \textbf{med} & $\bolds
{(Q_1$--$Q_3)}$ \\
\hline
1 & 260 & \phantom{0}0.0 & (0.0--0.0) & 256 & \phantom{0}0.0 &
(0.0--0.0) & 260 & \phantom{0}0.0 & (0.0--0.0) \\
2 & 248 & \phantom{0}6.1 & (5.9--6.7) & 241 & \phantom{0}6.1 &
(5.9--6.7) & 247 & \phantom{0}6.1 & (5.9--6.8) \\
3 & 226 & 12.2 & (11.8--12.9) & 224 & 12.2 & (11.8--12.9) & 224 & 12.2
& (11.8--12.9) \\
4 & 181 & 24.3 & (23.8--25.3) & 180 & 24.3 & (23.8--25.2) & 180 & 24.3
& (23.8--25.2) \\
5 & \phantom{00}3 & 23.4 & (23.4--26.5) & \phantom{00}2 & 23.4 &
(23.4--23.4) & \phantom{00}2 & 23.4 & (23.4--23.4) \\
\hline
\end{tabular*}  \vspace*{-3pt}
\end{table}

\subsection{Existing clustering methods, the need for extensions}\label{subsecexistingMethods}
In the literature numerous clustering methods applicable in many
different situations are available. Nevertheless,
as we discuss below, none of them is applicable for our problem of
clustering where each subject is represented
by a set of $R$, in general unbalanced and irregularly sampled
longitudinal profiles of markers which may have
a different nature starting from continuous and ending with a
dichotomous one.\vspace*{-2pt}

\subsubsection{Classical approaches and a mixture model-based
clustering}\label{subsubsecmodelBasedClust}
Apart from classical approaches
like hierarchical clustering or $K$-means method
[see, e.g., \citet{HastieTibshiraniFriedman09},\vadjust{\goodbreak} Chapter 13, \citet{JohnsonWichern07}, Chapter 12]
and their many extensions,
model-based clustering built on mixtures of parametric or even
nonparametrically specified distributions assumed
for random vectors $\Yb_1,\ldots,\Yb_N$ has become quite popular
in the past decade [e.g., \citet{FraleyRaftery02}]. This is probably
partially due to the availability of ready-to-use software like
R [\citet{Rbook12}] packages \texttt{mclust} of \citet{Mclust2006} for
clustering based on mixtures of multivariate normal distributions,
earlier versions of the \texttt{mixAK} package described by \citet{KomarekNMix}
or \texttt{mixtools} of \citet{Rmixtools09}, which also allows for
nonparametric estimation of the mixture components.
Another rapid evolution of model-based clustering algorithms also
originates from their need in gene-expression data analysis
[e.g., \citet{NewtonChung10,Witten11}].
\mbox{Nevertheless}, classical approaches, model-based clustering based on
mixtures of distributions and many other related methods
are not applicable in our context.

Some of the above mentioned methods rely on distances based on
a suitable metric between the observed values of underlying random vectors
$\Yb_1,\ldots,\Yb_N$
being viewed as points in the Euclidean space of a certain dimension.
However, in our situation the dimension of each $\Yb_i$ is generally
different for each subject and typically random. Hence,
it is even not possible to define a common sample space needed to
define a reasonable metric to calculate the distances.

For model-based methods, on the other hand,
it is necessary to assume that the random vectors $\Yb_1, \ldots, \Yb_N$
are independent and, given the classification, identically distributed
according to a suitable (multivariate) distribution.
Neither of these can be assumed since for typical longitudinal data
(including our {\PBCsub} data) the measurements are taken
at different time points for each subject and, hence, $\Yb_1, \ldots,
\Yb_N$ are hardly identically distributed even if the number of measurements
was the same for all subjects (which is also not the case for our data).

\subsubsection{Clustering based on a mixture of regression models}\label
{subsubsecmixtureRegression}
For data where the~$i$th subject out of $N$ to be classified may be
represented by one response random variable $Y_i$ and a vector of
possibly fixed
covariates $\xb_i$, several methods for clustering based on mixtures of
regression models have been developed. Among the first,
\citet{QuandtRamsey78} assume a two-component mixture of two normal
linear regressions. Extension into a general number of components
and also a practically applicable implementation is provided by \citet
{Rmixtools09}. A variant of the clustering based on a mixture of regressions
with application to gene-expression data is given by \citet{QinSelf06}.
A generalization, allowing also for nonnormally
distributed response random variables $Y_1,\ldots,Y_N$, is due to \citet
{GruenLeisch07}, who consider mixtures of generalized linear models.
To apply these methods for our application, the single response random
variables $Y_{i,r,j}$ could play the role of the response variables
in the mixtures of regression models and the time points $t_{i,r,j}$
the role of the fixed covariates.
Nevertheless, the clustering approaches based on mixtures
of regression models are also ruled out in our situation since (a) we
cannot assume a single parametric distribution for all response
variables in the data since both continuous and discrete response
variables appear in our data set,
(b) each subject is in general represented by more than one pair
(response, covariates).

\subsubsection{Clustering approaches for functional data and stochastic
processes}\label{subsubsec01clustFD}
For given $r \in\{1,\ldots,R\}$, a set $ \{\Yb_{1,r},\ldots,\Yb_{N,r}
\}$ of longitudinal trajectories of the $r$th marker
could also be viewed as a set of functional observations or, in more
general, a set of realizations of a certain random process.
A~similar setting is also found in the applications in genomics where
$\Yb_{i,r}$ is typically a vector representing
the expression curve of gene $i$ over time. In the functional data or
genomics literature, several clustering methods have been developed
for situations when it is possible to assume a decomposition of each
observed value into
%
\begin{equation}
\label{eq01decomp} Y_{i,r,j} = m_{i,r}(t_{i,r,j}) +
\varepsilon_{i,r,j},\qquad i=1,\ldots,N, j=1,\ldots,n_i,
\end{equation}
where $m_{i,r}(t)$ is either the value of the underlying random
functional or the mean $i$th gene-expression at time $t$
or, in general, the underlying stochastic process, and $\varepsilon
_{i,r,j}$ $(i=1,\ldots,N,$ $j=1,\ldots,n_{i,r})$
are random variables with a zero mean and either a common variance
$\sigma^2$ or subject/gene specific variances
$\sigma_i^2$ $(i=1,\ldots,N)$. Based on this model (\ref{eq01decomp}),
\citet{JamesSugar03} and \citet{LiuYang09} developed methods for the
clustering of functional data.
\citet{PengMueller08} proposed a distance-based clustering method and
apply it to data from online auctions.
For the genomics applications, \citet{RamoniSebastianiKohane02} present
an agglomerative clustering procedure
based on the autoregressive model in equation~(\ref{eq01decomp}).
Another gene-expression clustering application
based in fact on a mixture of regression models in equation (\ref
{eq01decomp}) is provided by \citet{Maet06}.
In our situation, these methods could only be applied if there is only
one continuous marker available for each patient.
Hence, with the {\PBCsub} data, clustering would have to be based only
on either {\lbili} values or {\platelet} values
if it was assumed that they come from a continuous location-shift
distribution. The dichotomous {\spiders} values
cannot be used at all.

\subsubsection{Clustering based on mixture extensions of the mixed
models}\label{subsubsecMixtureMixed}
For the analysis of the continuous longitudinal data, the linear mixed
model [LMM, \citet{LairdWare82}] plays
a prominent role. For given $r \in\{1,\ldots,R\}$, it is based on expression
%
\begin{equation}
\label{eq01LMM} \Yb_{i,r} = \tra{\xb_{i,r}}
\alphab_r + \tra{\zb_{i,r}}\bb_{i,r} +
\epsb_{i,r},\qquad i=1,\ldots,N,
\end{equation}
where $\xb_{i,r}$ and $\zb_{i,r}$ are vectors of fixed covariates
containing the time points $t_{i,r,j}$ $(j=1,\ldots,n_{i,r})$
and possibly other factors.
Further, $\alphab_r$ is a vector of unknown regression parameters,
$\bb_{i,r}$ are i.i.d. random variables---random effects with unknown
mean $\betab_r$ and a covariance matrix $\Db_r$,
and $\epsb_{i,r}$ are independent random vectors with zero mean and
a covariance matrix $\Sigmab_{i,r}$.
To cluster subjects based on the continuous longitudinal data, several
approaches stemming from a mixture extension of the LMM (\ref{eq01LMM})
have been proposed in the literature. \citet{VerbekeLesaffre96} assume
a normal mixture in the distribution of random effects
and apply their method to clustering of growth curves,
while \citet{CeleuxMartinLavergne05} consider a mixture of linear
mixed models and perform clustering of gene-expression data.
\citet{CruzMesiaQuintanaMarshall08} proceed in a similar way, however,
they replace the $\tra{\xb_{i,r}}\alphab_r + \tra{\zb_{i,r}}\bb_{i,r}$
part of (\ref{eq01LMM}) by a nonlinear expression in $\alphab_r$ and
$\bb_{i,r}$.

By a suitable choice of the covariate vectors and imposing a suitable
structure on the error covariance matrices $\Sigmab_i$,
it is possible to use the LMM (\ref{eq01LMM}) also for the analysis of
$R > 1$ continuous longitudinal markers
and for clustering based on it as was done by \citet
{VillarroelMarshallBaron09},
or could be done using the model of \citet{KomarekHansenet10} who
performed the discriminant analysis, though.
However, analogously to Section~\ref{subsubsec01clustFD}, all
mentioned methods could be used for our application only
if we wanted to base the clustering only on {\lbili} and/or {\platelet} values.

One possible strategy for clustering based on not only continuous
longitudinal profiles is to use a general form of the model
proposed by \citet{BoothCasellaHobert08} [equation (3) in their paper],
where they assume that the (not necessarily normal) distribution of longitudinal
observations of a particular marker depends on cluster-specific
parameters and on a vector of random effects. Nevertheless,
except for this general definition, they focus in their paper on
a linear mixed model which is only applicable in situations
when the observed longitudinal markers are continuous.

A specific option which allows for clustering based on a single
longitudinal marker of a discrete nature is to replace the LMM (\ref
{eq01LMM}) by
a generalized linear mixed model [GLMM, e.g., \citet
{MolenberghsVerbeke05}] and assume a suitable mixture
in the distribution of random effects [\citet
{SpiessensVerbekeKomarek02}]. An example of clustering based on such a model
is shown in \citet{MolenberghsVerbeke05}, Section 23.2. Nevertheless,
it is still not possible to jointly use all three (in general, all $R$) markers.

\subsubsection{Objectives and outline of the paper}\label{subsubsecObjectives}
In previous paragraphs we gave a brief overview of the most common
classes of clustering approaches. We also argued that
none of them are capable of exploiting jointly irregular longitudinal
measurements of $R\geq1$ markers of different nature
(continuous, discrete or even dichotomous) as it is required by the
{\PBCsub} data. Even though our overview
is by no means exhaustive, we are not aware of any method that would
meet such needs. For these reasons we propose
a clustering method that will be built upon the multivariate extension
(where the word ``multivariate'' points to the fact that $R \geq1$
markers will be modeled jointly)
of the GLMM with a normal mixture in the distribution of random effects
proposed by \citet{SpiessensVerbekeKomarek02}, [SVK], and \citet{MolenberghsVerbeke05}, [MV].
Not only did they model just the $R=1$ longitudinal outcome,
but they also considered only a homoscedastic normal mixture.
Nevertheless, this is a rather restrictive assumption, especially in
our context where
each mixture component should represent one cluster.
Hence, to have a better ground for clustering, the heteroscedastic
mixture will be considered in our proposal.
Further, in their illustrations, [SVK] and [MV]
usually included at most bivariate random effects. This was probably
due to the fact that
as a method of estimation they exploited the maximum-likelihood through
the EM algorithm which starts to be
computationally troublesome for models with random effects of a higher
dimension.
For computational complexity implied by the multivariate extension of
the mixture GLMM,
but not only because of this, we shall use the Bayesian inference based
on the Markov chain Monte Carlo (MCMC) simulation here.

In Section~\ref{secMMGLMM} we next describe the multivariate mixture
generalized linear mixed model which will serve
as the basis for our clustering procedure and show how to apply it to
the {\PBCsub} data. The clustering procedure will be described,
and the clustering of patients from the {\PBCsub} data will be
performed in Section~\ref{secclustering}.
In Section~\ref{secselectK} we discuss the possibility of estimating
a number of clusters needed in situations
when this does not follow from the context.
We evaluate the proposed methodology in Section~\ref{secSimul} on
a simulation study and finalize the paper
by a discussion in Section~\ref{secDiscuss}.


\section{Mixture multivariate generalized linear mixed model}\label{secMMGLMM}

\subsection{Model specification}
Our proposed clustering procedure is based on a multivariate mixture
generalized linear mixed model (MMGLMM).
We first express the conditional mean of each response profile using
a standard GLMM, that is,
%
\begin{eqnarray}
\label{eqGLMM} h_r^{-1} \bigl\{\EV(Y_{i,r,j} |
\alphab_r, \bb_{i,r}) \bigr\} = \tra{\xb_{i,r,j}}
\alphab_r + \tra{\zb_{i,r,j}}\bb_{i,r},
\nonumber
\\[-8pt]
\\[-8pt]
\eqntext{i=1,\ldots,N, r=1,\ldots,R, j=1,\ldots,n_{i,r},}
\end{eqnarray}
where $h_r^{-1}$ is the link function used to model the mean of the
$r$th marker,
$\xb_{i,r,j}$, $\zb_{i,r,j}$ are
vectors of known covariates which may include a constant for intercept,
time values
in which the longitudinal observations have been taken
or any other additional covariates.
Further, $\alphab_r$ is
a vector of unknown regression coefficients (fixed effects)
and $\bb_{i,r}$ is
a vector of random effects for the $r$th response specific for the
$i$th subject.
We assume hierarchically centered GLMM [\citet{Gelfandet95}] where the
random effects
$\bb_{1,r},\ldots,\bb_{N,r}$ have in general a nonzero and unknown
mean, let's say $\betab_r$, $r=1,\ldots,R$;
see equation (\ref{eqbeta}) below.
Being within the GLMM framework, we assume that for each $i=1,\ldots
,N$, $r=1,\ldots,R$, $j=1,\ldots,n_{i,r}$,
the conditional distribution
$p(y_{i,r,j} | \phi_r, \alphab_r, \bb_{i,r})$ belongs to an exponential
family with the mean
specified by (\ref{eqGLMM}),\vspace*{1pt} and possibly unknown dispersion parameter
$\phi_r$.
In a sequel, let $\psib= (\tra{\phib}, \tra{\alphab}\tra{ )},$ where
$\alphab= (\tra{\alphab_1},\ldots,\tra{\alphab_R}\tra{ )}$,
$\phib= (\tra{\phi_1},\ldots,\tra{\phi_R}\tra{ )}$ is the vector of
GLMM related parameters.

Further, let $\bb_i = (\tra{\bb_{i,1}},\ldots,\tra{\bb_{i,R}}\tra{)}$
be
a joint vector of random effects for the $i$th subject $(i=1,\ldots,N)$.
Dependence between the $R$ longitudinal markers of a particular subject
$i$ represented by the response vectors
$\Yb_{i,1}$, $\ldots,$ $\Yb_{i,R}$ is taken into account by assuming
a joint distribution for the random effect vector
$\bb_i$ which also grounds our clustering procedure. We assume that the
$i$th subject
belongs to one of a fixed number of $K$ clusters (see Section \ref
{secselectK} for possible approaches to choose~$K$
if this does not follow from the context of the application at hand),
each cluster with a probability
$w_k = \Pr(u_i = k | \wb)$, $(0 \leq w_k \leq1$, $k=1,\ldots,K$, $\sum
_{k=1}^K w_k = 1)$,
where $u_i \in \{1,\ldots,K \}$ is the $i$th subject allocation and
$\wb= (w_1,\ldots,w_K\tra{)}$.
We further assume that the corresponding random effect vector $\bb_i$
follows a multivariate normal distribution with an unknown mean $\mub_{u_i}$
and a (generally nondiagonal) unknown covariance matrix $\Db_{u_i}$,
that is,
\[
p(\bb_i | \thetab, u_i = k) = \varphi(
\bb_i; \mub_k, \Db_k ),\qquad i=1,\ldots,N, k=1,
\ldots,K,
\]
where $\varphi(\cdot| \mub, \Db)$ is a density of the (multivariate)
normal distribution
with a mean $\mub$ and a covariance matrix $\Db$,
and $\thetab= (\tra{\wb}, $ $\tra{\mub_1},\ldots,\tra{\mub_K}, $
$\vec(\Db_1),\ldots,\break\vec(\Db_K)\tra{ )}$ is a vector of unknown
parameters related to the
distribution of random effects.
That is, overall,
we assume a multivariate normal mixture in the distribution of random effects:
%
\begin{equation}
\label{eqdistb} \bb_i | \thetab\iid\sum
_{k=1}^K w_k \MVN(\mub_k,
\Db_k ),\qquad i=1,\ldots,N.
\end{equation}
With this approach, we represent the unbalanced longitudinal observations
$\Yb_1,\ldots,\Yb_N$ using a set of i.i.d.
random vectors $\bb_1,\ldots,\bb_N$ which allows us to develop
a clustering procedure based on ideas
of the mixture model-based clustering introduced in Section~\ref
{subsubsecmodelBasedClust}.

Finally, we point out that given our model, the mean effect (in a total
population) of covariates included in
the vectors $\zb_{i,r,j},$ $i=1,\ldots,N,$ $r=1,\ldots,R,$ $j=1,\ldots
,n_{i,r}$ is given by
%
\begin{equation}
\label{eqbeta} \betab= \sum_{k=1}^K
w_k\mub_k.
\end{equation}
This is a vector composed of $R$ subvectors, say, $\betab_1,\ldots
,\betab_R$, which play the role
of the fixed effects for covariates included in the $z$-covariate vectors.
Hence, for identifiability reasons, it is assumed that the
vectors $\xb_{i,r,j}$ and $\zb_{i,r,j},$ $i=1,\ldots,N,$ $r=1,\ldots
,R,$ $j=1,\ldots,n_{i,r}$ do not contain the same covariates.

\subsection{Likelihood, Bayesian estimation}\label{subsecLikBayes}
For largely computational reasons, we shall use the Bayesian inference based
on the output from the MCMC simulation.
To this end, the model must be specified also from a Bayesian point of view.
A~priori, we assume the independence between the mixture related
parameters $\thetab$ and the GLMM related parameters $\psib$.
That is, the prior distribution $p(\psib, \thetab)$ factorizes as
$p(\psib, \thetab) = p(\psib) \times p(\thetab)$.
The factorization of the prior is typical in generalized linear mixed models
and might be justified in our application because the parameters involved
in the $\psib$ and $\thetab$ vector, respectively, express different
features of the model.
Specifically, for $p(\thetab)$, we use a multivariate version of the
classical proposal of \citet{RichardsonGreen97}
as prior distribution, and, for $p(\psib)$, we adopt classically used
priors in this context
[see, e.g., \citet{FongRueWakefield10}].
A~detailed description of the assumed form of $p(\psib, \thetab)$
is provided in Appendix A 
of the Supplementary Material [\citet{glmmClustWeb}] where we also explain how to choose
the prior hyperparameters in order to obtain a weakly informative prior
distribution.

The likelihood of the MMGLMM follows from (\ref{eqGLMM}) and (\ref{eqdistb}):
%
\begin{equation}
\label{eqLobs} L(\psib, \thetab) = p(\yb| \psib, \thetab) = \prod
_{i=1}^N \Biggl(\sum_{k=1}^K
w_k L_{i,k}(\psib, \thetab) \Biggr),
\end{equation}
where
%
\begin{eqnarray}
\label{eqLik}
L_{i,k}(\psib,
\thetab) = \int\Biggl\{ \prod_{r=1}^R\prod
_{j=1}^{n_{i,r}}p(y_{i,r,j} |
\phi_r, \alphab_r, \bb_{i,r}) \Biggr\} p(
\bb_i | \thetab, u_i=k)\,d\bb_{i},
\nonumber
\\[-8pt]
\\[-8pt]
\eqntext{i=1,\ldots,N, k=1,\ldots,K}
\end{eqnarray}
is the contribution of the $i$th subject to the likelihood under the
assumption that the random effects
are distributed according to the $k$th mixture component.

%
\begin{figure}[b]
\vspace*{-3pt}
\includegraphics{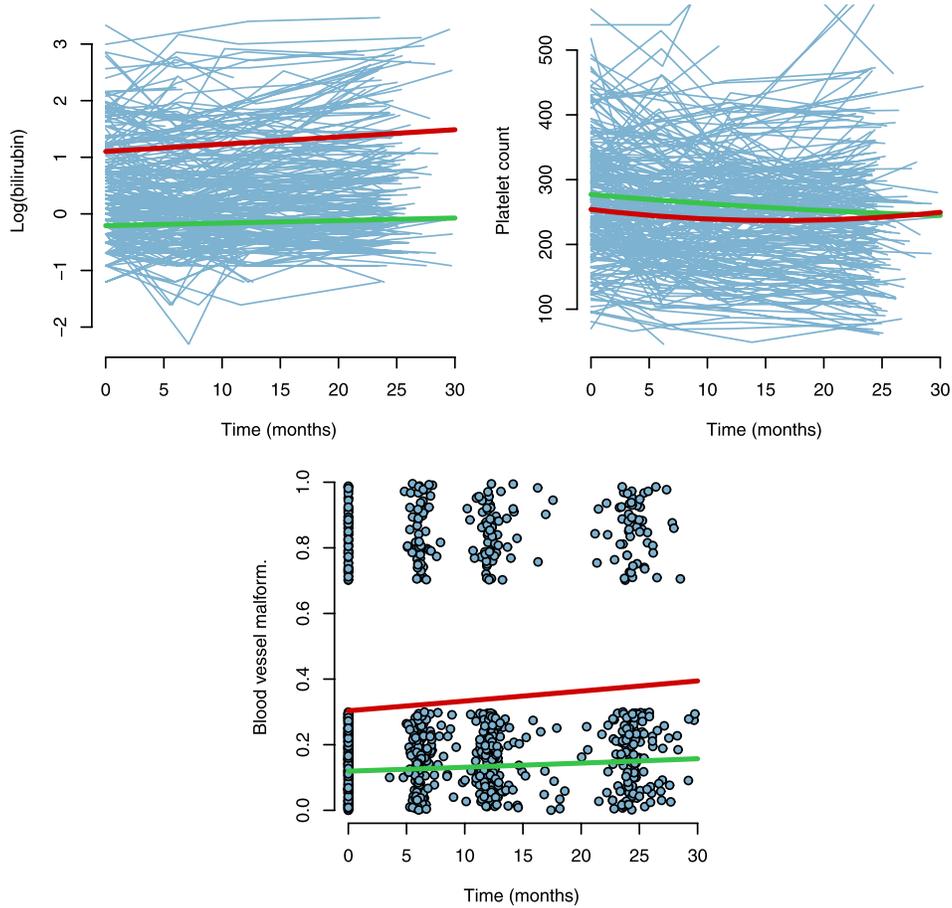}

\caption{{\PBCsub} data. Observed values of the longitudinal markers.
Thick lines show cluster-specific marginal mean evolution over time
based on posterior means
of the mixture means $\mub_1$ (green) and $\mub_2$ (red).
Observed values of dichotomous blood vessel malformations (\spiders)
are vertically jittered.}
\label{figPBC01}
\end{figure}

MCMC methods are used to generate a sample
$\Samp_M =
\{(\psib^{(m)}, \thetab^{(m)}): m=1,\ldots,M \}$
from the posterior distribution
$p (\psib, \thetab | \yb) \propto L(\psib, \thetab) \times p(\psib,
\thetab).$
Namely, a block Gibbs algorithm is used with the Metropolis--Hastings
steps for those blocks of model
parameters where the normalizing constant of the full conditional
distribution does not have a closed form.
A well-known identifiability problem which arises from the invariance
of the likelihood
under permutation of the component labels is solved by applying the relabeling
algorithm of \citet{Stephens00},\vadjust{\goodbreak} which is suitable for mixture models
targeted toward clustering in particular.
For details of the MCMC algorithm, refer to Appendix B 
of the Supplement [\citet{glmmClustWeb}].

\subsection{MMGLMM for the PBC910 data}\label{subsecPBCMMGLMM}
The MMGLMM for the clustering of patients included in the PBC910 data
will be based on longitudinal measurements of
(1) logarithmic serum bilirubin (\textsf{lbili}, $Y_{i,1,j}$),
(2) platelet counts (\textsf{platelet}, $Y_{i,2,j}$)
and (3) dichotomous presence of blood vessel malformations (\textsf
{spiders}, $Y_{i,3,j}$)
with assumed (1) Gaussian, (2) Poisson and (3) Bernoulli distribution,
respectively.
Exploration of the observed longitudinal profiles (see also Figure\vadjust{\goodbreak} \ref
{figPBC01})
suggests the following form of the mean structure~(\ref{eqGLMM}):
%
\begin{eqnarray}\label{eqmodelPBC}
\EV(Y_{i,1,j} | \bb_{i,1}) &=&
b_{i,1,1} + b_{i,1,2} t_{i,1,j},
\nonumber\\
\log\bigl\{\EV(Y_{i,2,j} | \bb_{i,2}) \bigr\} &=&
b_{i,2,1} + b_{i,2,2} t_{i,2,j},
\\
\logit\bigl\{\EV(Y_{i,3,j} | b_{i,3}, \alpha_3)
\bigr\} &=& b_{i,3} + \alpha_3 t_{i,3,j},\nonumber
\end{eqnarray}
$i=1,\ldots,N$, $j=1,\ldots,n_{i,r}$, $r=1, 2, 3$, where $1 \leq
n_{i,r} \leq5$.
In model (\ref{eqmodelPBC}), $t_{i,r,j}$ is the time in months from the
start of the follow-up when the value of $Y_{i,r,j}$
was obtained.

In the main analysis, we will classify patients into two groups and,
hence, a~two component mixture ($K=2$)
will be considered in the distribution of five-dimensional random
effect vector
$\bb_i = (b_{i,1,1},b_{i,1,2},$ $b_{i,2,1},b_{i,2,2},$ $b_{i,3}\tra{
)}$, where $b_{i,1,1}, b_{i,2,1}, b_{i,3}$
are random intercepts from the GLMM for each marker and $b_{i,1,2},
b_{i,2,2}$ are random slopes from the GLMM for the first two markers.
The model also involves the fixed effect $\alphab= \alpha_3$, the slope
from the logit model for the third Bernoulli response and a dispersion parameter
$\phi_1 = \var(Y_{i,1,j} | \bb_{i,1})$, the residual variance
from the Gaussian model for the first marker, the logarithmic bilirubin.
Let $\sigma_1 = \sqrt{\phi_1}$ be the corresponding residual standard deviation.
The GLMM related parameters are thus $\psib= (\sigma_1, \alpha_3\tra{ )}$.
The results that we report are based on 10,000 iterations of 1:100
thinned MCMC obtained after a burn-in period
of 1000 iterations.
See Appendix C 
of the Supplement [\citet{glmmClustWeb}]
for a full Bayesian specification of the model, particular choices of
the hyperparameters, an illustration of the performance
of the MCMC and detailed results.\looseness=-1

With respect to clustering, the most important parameters are the
mixture weights $w_1$, $w_2$
and the mixture means $\mub_1$, $\mub_2$ which characterize the clusters.
Their estimates taken to be the posterior means
(denoted by $\widehat{w}_1, \widehat{w}_2$, $\widehat{\mub}_1, \widehat
{\mub}_2$,
resp.) estimated from an appropriately relabeled MCMC sample are given
in Table~\ref{tabPBC03}
together with the 95\% highest posterior density credible intervals
(HPD CI).
The first cluster is thus characterized by a remarkably lower baseline
bilirubin level and its slower increase over time
compared to the second cluster. For the platelet counts, there is
almost no difference between the clusters
at the baseline and only a moderate difference with respect to the rate\vadjust{\goodbreak}
of its change, with the second cluster
showing a faster decline. Finally, blood vessel malformations exhibit
a higher probability in the second cluster
compared to the first one. From the clinical point of view, the first
cluster exhibits more favorable
values and also an evolution of all three markers and, hence, it should
correspond to patients with
a better prognosis compared to the second cluster. We confirm this
conclusion in Section~\ref{subsecclusPBC}
upon the classification of the individual patients.

%
\begin{table}
\caption{{\PBCsub} data. Posterior means and 95\% HPD credible
intervals for mixture weights, mixture means
and GLMM related parameters}
\label{tabPBC03}
\begin{tabular*}{\textwidth}{@{\extracolsep{\fill}}lcc@{}}
\hline
& \multicolumn{1}{c}{$\bolds{k = 1}$} & \multicolumn{1}{c@{}}{$\bolds{k
= 2}$}\\[-4pt]
& \multicolumn{1}{c}{\hrulefill} & \multicolumn{1}{c@{}}{\hrulefill}\\
& \multicolumn{1}{c}{$\bolds{\widehat{w}_1 = \EV(w_1 | \yb) = 0.598}$} &
\multicolumn{1}{c}{$\bolds{\widehat{w}_2 = \EV(w_2 | \yb) = 0.402}$}\\
\textbf{Parameter} & \multicolumn{1}{c}{$\textbf{(0.471, 0.711)}$} &
\multicolumn{1}{c}{$\textbf{(0.289, 0.529)}$}\\
\hline
\multicolumn{3}{c}{\textit{Logarithmic bilirubin \textup{(}\lbili\textup{)}}} \\
Intercept & $-$0.209 & 1.102\\
$\widehat{\mu}_{k,1} = \EV(\mu_{k,1} | \yb)$ & ($-$0.332, $-$0.082) & (0.828,
1.387)\\[3pt]
Slope & 0.00450 & 0.01281\\
$\widehat{\mu}_{k,2} = \EV(\mu_{k,2} | \yb)$ & (0.00056, 0.00818) & (0.00476,
0.02108)\\[3pt]
Residual std. dev. & \multicolumn{2}{c}{0.314} \\
$\widehat{\sigma}_1 = \EV(\sigma_{1} | \yb)$ & \multicolumn
{2}{c}{(0.294, 0.333)}
\\[6pt]
\multicolumn{3}{c}{\textit{Platelet count \textup{(}\platelet\textup{)}}} \\
Intercept & 5.58 & 5.46\\
$\widehat{\mu}_{k,3} = \EV(\mu_{k,3} | \yb)$ & (5.49, 5.65) & (5.35,
5.58)\\[3pt]
Slope & $-$0.00567 & $-$0.00828\\
$\widehat{\mu}_{k,4} = \EV(\mu_{k,4} | \yb)$ & ($-$0.00799,
$-$0.00339) & ($-$0.01354,
$-$0.00306)\\[3pt]
\multicolumn{3}{c}{\textit{Presence of blood vessel malformations
\textup{(}\spiders\textup{)}}} \\
Intercept & $-$4.33 & $-$0.83\\
$\widehat{\mu}_{k,5} = \EV(\mu_{k,5} | \yb)$ & ($-$5.90, $-$2.88) & ($-$1.66,
$-$0.02)\\[3pt]
Slope & \multicolumn{2}{c}{0.0280} \\
$\widehat{\alpha}_3 = \EV(\alpha_{3} | \yb)$ & \multicolumn
{2}{c}{(0.0026, 0.0532)} \\
\hline
\end{tabular*}        \vspace*{-3pt}
\end{table}

%
\begin{table}
\caption{{\PBCsub} data. Standard deviations (on a diagonal) and
correlations (off-diagonal elements)
for each mixture component derived from the posterior means $\widehat
{\Db}_1 = \EV({\Db}_{1} | \yb)$
and $\widehat{\Db}_2 = \EV({\Db}_{2} | \yb)$ of the mixture covariance matrices}
\label{tabPBC05}
\begin{tabular*}{\textwidth}{@{\extracolsep{\fill
}}lcd{2.5}d{2.3}d{2.4}d{2.3}@{}}
\hline
& \multicolumn{1}{c}{\textbf{Intercept}} & \multicolumn{1}{c}{\textbf
{Slope}} & \multicolumn{1}{c}{\textbf{Intercept}} &
\multicolumn{1}{c}{\textbf{Slope}} & \multicolumn{1}{c@{}}{\textbf
{Intercept}} \\
& \multicolumn{1}{c}{\textbf{(lbili)}} & \multicolumn{1}{c}{\textbf
{(lbili)}} & \multicolumn{1}{c}{\textbf{(platelet)}} &
\multicolumn{1}{c}{\textbf{(platelet)}} & \multicolumn{1}{c@{}}{\textbf
{(spiders)}}\\
\hline
& \multicolumn{5}{c}{$k = 1$} \\
Intercept (lbili) & 0.428 & 0.031 & -0.282 & -0.086 & 0.326\\
Slope (lbili) & & 0.00837 & 0.040 & -0.214 & 0.100\\
Intercept (platelet) & & & 0.309 & -0.039 & -0.042\\
Slope (platelet) & & & & 0.0105 & 0.028\\
Intercept (spiders) & & & & & 4.02\\[3pt]
& \multicolumn{5}{c}{$k = 2$} \\
Intercept (lbili) & 0.776 & -0.183 & 0.119 & -0.139 & 0.171\\
Slope (lbili) & & 0.03090 & -0.034 & 0.249 & 0.116\\
Intercept (platelet) & & & 0.398 & -0.046 & -0.191\\
Slope (platelet) & & & & 0.0232 & -0.043\\
Intercept (spiders) & & & & & 2.42\\
\hline
\end{tabular*}
\end{table}

To get a better idea of the meaning of the clusters, we used
$\widehat{\mub}_1$ and $\widehat{\mub}_2$ together with the posterior
means of the GLMM related parameters
$\psib$ (see Table~\ref{tabPBC03}) and the posterior means of the
mixture covariance matrices $\Db_1$ and $\Db_2$
(see Table~\ref{tabPBC05}),\vadjust{\goodbreak} and calculated the estimates of the cluster
specific (marginal) mean longitudinal
evolutions $\EV(Y_{\cdot,r,\cdot} | \alphab_r, u=k) = \EV_{\bb} \{\EV
(Y_{\cdot,r,\cdot} | \bb, \alphab_r, u=k) \}$,
$k=1,2,$ $r=1,2,3$ over time. These are plotted as green ($k=1$) and
red ($k=2$) lines on Figure~\ref{figPBC01}.

\section{Clustering procedure}\label{secclustering}
It follows from the decision theory for classification [see \citet
{HastieTibshiraniFriedman09}, Section 2.4]
that the optimal classification of the $i$th subject ($i=1,\ldots,N$)
is to be based on the posterior component probabilities
$\pi_{i,k} = \Pr(u_i = k | \yb)$ $(k=1,\ldots,K)$. In our case, they
are calculated by marginalization
over the posterior distribution as
%
\begin{eqnarray}
\label{eqpiik}
\pi_{i,k} & = &\int p_{i,k}(
\psib, \thetab) p (\psib, \thetab| \yb)\,d(\psib, \thetab) = \EV\bigl
\{p_{i,k}(\psib, \thetab) | \yb\bigr\}
\nonumber
\\[-10pt]
\\[-10pt]
\nonumber
& \approx&\frac{1}{M}\sum_{m=1}^M
p_{i,k} \bigl(\psib^{(m)}, \thetab^{(m)} \bigr) =
\widehat{\pi}_{i,k},
\end{eqnarray}
where by Bayes' rule
%
\begin{eqnarray}
\label{eqpik}
p_{i,k}(\psib, \thetab) = \Pr
(u_i = k | \psib, \thetab, \yb_i ) = \frac{w_k L_{i,k}(\psib, \thetab
)}{\sum_{l=1}^K w_l L_{i,l}(\psib, \thetab)},
\nonumber
\\[-10pt]
\\[-10pt]
\eqntext{i=1,\ldots,N, k=1,\ldots,K.}
\end{eqnarray}
The $i$th subject is classically assigned to the cluster $g_i$ for
which $\widehat{\pi}_{i,g_i}$ is largest among (\ref{eqpiik})
[e.g., \citet{Titteringtonet85,McLachlanBasford88}].

In non-Bayesian applications, the clustering procedure is usually based
on the values of
$\widehat{p}_{i,k} = p_{i,k}(\widehat{\psib}, \widehat{\thetab})$,
where $\widehat{\psib}$ and $\widehat{\thetab}$ are suitable estimates
(e.g., maximum-likelihood) of the model parameters.
Only rarely the uncertainty in the estimation of $\widehat{p}_{i,k}$
expressed by evaluating their standard errors or
calculating the confidence intervals is taken into account. This is
probably because of the fact that $p_{i,k}(\psib, \thetab)$
depend on the model parameters in a relatively complex way. In our case
of the MMGLMM, it is, for example, complicated by the necessity to
integrate the GLM likelihood over the assumed distribution of the
random effects [see equation (\ref{eqLik})],
which in general does not have an analytic solution.

With the Bayesian approach followed in this paper, the values of
$\widehat{\pi}_{i,k}$ used for clustering
are the estimated posterior means of $p_{i,k}(\psib, \thetab)$ and with
the MCMC-based posterior inference,
we can easily evaluate also the posterior standard deviations
(counterparts of the classical standard errors)
or the credible intervals (counterparts of the classical confidence
intervals). These can be used to incorporate
an uncertainty in the classification which, for example, in the
clinical setting of the {\PBCsub} data,
can serve for the identification of subjects that should undergo
additional screening before their ultimate classification;
see Section~\ref{subsecclusPBC}.

\subsection{Clustering of patients from the PBC910 data}\label{subsecclusPBC}
\begin{figure}

\includegraphics{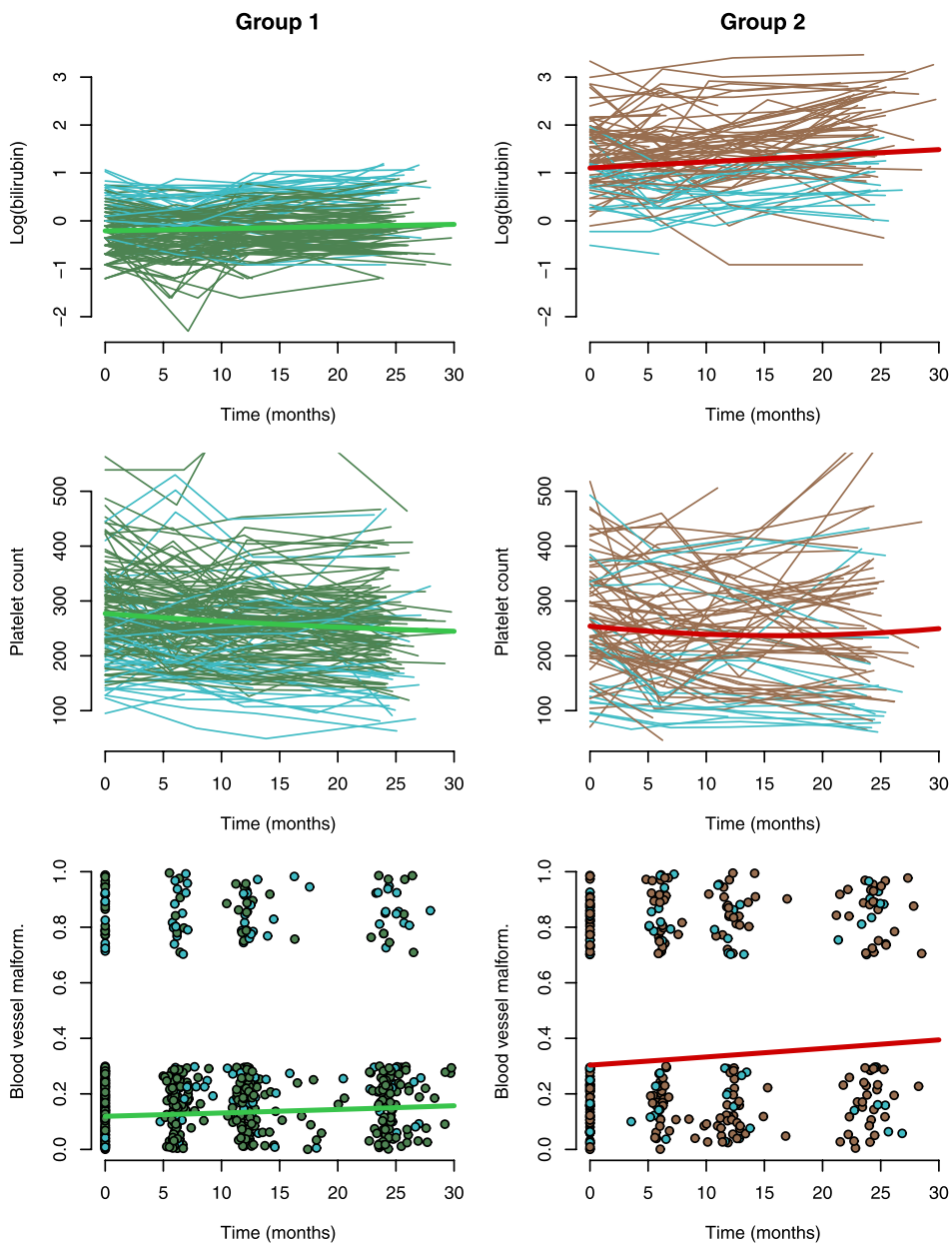}

\caption{{\PBCsub} data. Observed values of the longitudinal markers
upon classification into $K=2$ groups.
The thick lines show cluster-specific marginal mean
evolution over time based on posterior means of the mixture means $\mub
_1$ (green) and $\mub_2$ (red).
Observed values of dichotomous blood vessel malformations (\spiders)
are vertically jittered.
Profiles of patients for whom the lower limit of the 95\% HPD credible
interval did not exceed
0.5 are drawn in light blue.}\label{figPBC02}
\end{figure}

\begin{figure}

\includegraphics{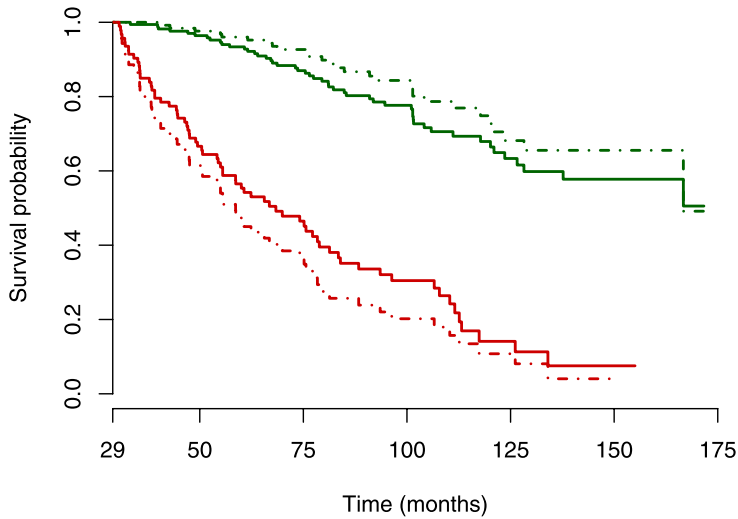}

\caption{{\PBCsub} data. Kaplan--Meier estimates of survival
probability beyond 910 days in each group ($k=1$: green, $k=2$: red)
created using the clustering procedure. Solid lines: everybody
classified using the maximal value of $\widehat{\pi}_{i,k}$,
dotted--dashed line: only patients for whom the lower limit of the 95\%
HPD CI for the component probability
exceeded 0.5 were classified.}
\label{figPBC10}
\end{figure}

Classification of patients according to the maximal value of the
estimated posterior mean $\widehat{\pi}_{i,k}$ leads
to 167 patients being classified in group 1 and 93 patients in group 2;
see Figure~\ref{figPBC02}.
We argued in Section~\ref{subsecPBCMMGLMM} that from a clinical point
of view, the first group should correspond
to patients with a better prognosis compared to the second group. With
the {\PBCsub} data, it is possible to confirm this conclusion
since the information concerning the residual progression free survival
time, defined as time till death due to liver
complications or till liver transplantation, is available in the form
of the classical right-censored data.
We calculated Kaplan--Meier estimates of the survival probabilities
based on data from patients classified in each group.
These are plotted as solid lines on Figure~\ref{figPBC10}. Indeed, the
survival prognosis of group 1 is much better than that
of group 2 with the estimated 5-year survival probability in group~1 of
0.934 compared to 0.554
in group 2, and the 10-year survival probabilities 0.679 and 0.141 in
groups 1 and 2, respectively.

%

%

%
\begin{figure}

\includegraphics{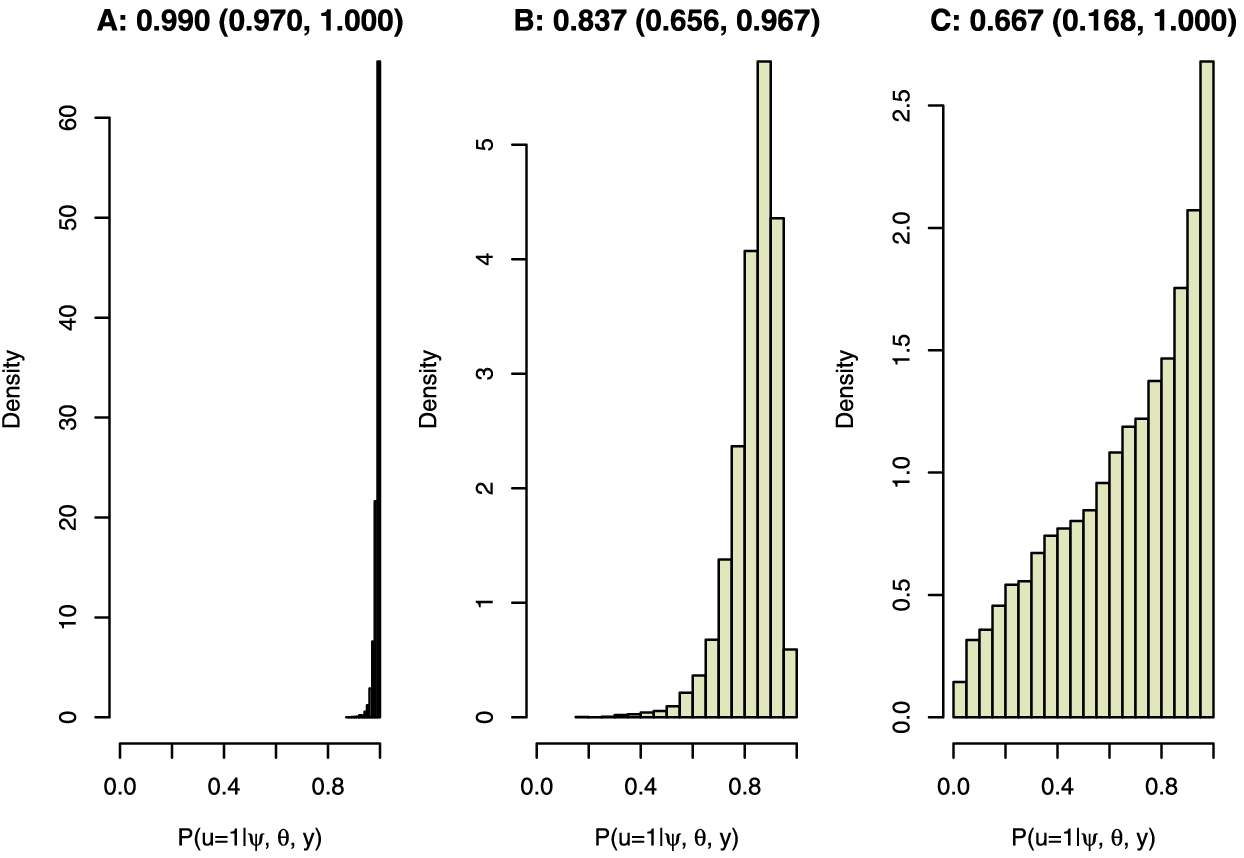}

\caption{{\PBCsub} data. Histograms of sampled values of component
probabilities $p_{i,1}(\psib, \thetab)$ for three selected patients.
Above the plot: estimated posterior mean $\widehat{\pi}_{i,1}$ and the
95\% HPD CI for $p_{i,1} = p_{i,1}(\psib, \thetab)$.}
\label{figPBC09}
\end{figure}

The fact that the posterior means $\pi_{i,k}$ of the patient specific
component probabilities characterize with
different certainty the probabilities of the allocation for different
patients is illustrated by Figure~\ref{figPBC09}.
It shows the MCMC-based estimates of the posterior distributions of the
first component probabilities
$p_{i,1} = p_{i,1}(\psib, \thetab)$ for three selected patients who
were all classified in a better prognosis group 1.
For patient A, $\widehat{\pi}_{i,1} = 0.990$ with a very narrow 95\%
HPD CI of $(0.970, 1.000)$ and, thus, her classification
in group 1 is almost certain. This is further confirmed by her
progression-free survival time, which is almost 14 years.
The posterior probability of belonging to group 1 is lower for
patient C ($\widehat{\pi}_{i,1} = 0.667$).
Nevertheless, it is still twice as high as the posterior probability of
belonging to group 2 ($\widehat{\pi}_{i,2} = 0.333$)
and, hence, it seems that patient C also belongs most likely in
group 1. On the other hand, his progression-free survival
time is only 3 years and 4 months (i.e., only 10 months beyond the time
point at which we perform classification)
and, hence, from a clinical point of view, this patient should rather
be classified in group 2. In this case, the uncertainty
in classification is expressed by a very wide 95\% HPD CI which is
$(0.168, 1.000),$ covering the majority
of the interval $(0, 1)$.

In clinical practice, the diagnostic procedure usually proceeds in
several steps, where in each step it is decided
whether it is possible to classify a patient with enough certainty or
whether additional examinations
are needed before the ultimate classification is determined. Quite
naturally, one of the diagnostic steps
of such a procedure could be based on calculated credible intervals for
individual component probabilities
$p_{i,k}=p_{i,k}(\psib, \thetab)$. The patient would be ultimately
classified in one of the considered groups
only if the lower limit of the corresponding
credible interval exceeds a certain threshold, say, $0.5$ in the
simplest case of classification into $K=2$ groups.
Applying this procedure to the {\PBCsub} data lead to 126 patients
being classified in group 1 and 70 in group 2.
In total, 64 patients (41 and 23 originally classified in group 1 and
group 2, respectively,
see light blue profiles on Figure~\ref{figPBC02}) remain without
ultimate classification, and additional
screening or examinations would have been recommended for them before
making a final decision.
The fact that the two groups consisting of only 126 $+$ 70 ultimately
classified patients better reflect
the progression free survival status is illustrated by the
Kaplan--Meier survival curves (dotted--dashed lines on Figure~\ref{figPBC10}),
which are now faster, diverting with the estimated 5-year survival
probability in group 1 of 0.960 compared to 0.465
in group 2, and the 10-year survival probabilities 0.748 and 0.108 in
groups 1 and 2, respectively.

\section{Selection of a number of clusters}\label{secselectK}
Until now, we assumed that the number of clusters, $K$, was known in
advance and fixed.
In medical applications where the found clusters are expected to
correspond to
certain prognostic groups, this is quite a reasonable assumption. Nevertheless,
in many other situations, the number of clusters should rather be
inferred from the data themselves.

With our approach, the selection of a number of clusters corresponds to
the selection of a number of mixture components
in the underlying distribution of the random effects of the MMGLMM. This
can also be viewed as a problem of model selection or model comparison.
Nevertheless,
as described, for example, in \citet{McLachlanPeel00}, Chapter~6, the
model comparison in the mixture setting
is complicated by the fact that the classical regularity conditions do
not hold. For this reason,
use of various sorts of information criteria is predominantly preferred
to classical testing
in most frequentist applications of the mixture models. In particular,
the Bayesian information criterion
[BIC, \citet{Schwarz78}] proved to be a useful tool for selecting the
number of mixture components
[\citet{DasguptaRaftery98}, \citet{FraleyRaftery02}, \citet{Hennig04},
\citet{CruzMesiaQuintanaMarshall08} and many others].

In Bayesian statistics, the Bayes factors [\citet{KassRaftery95}], for
which the BIC is an approximation, are widely recognized
as a tool of model selection. However, as pointed out by \citet
{Plummer08}, the Bayes factors have some practical limitations.
First, they cannot be routinely calculated from the MCMC output.
Second, they are numerically unstable when proper, but weakly
informative diffused priors (as it is in our case) are used.
In Bayesian applications, the deviance information criterion [DIC,
\citet{Spiegelhalteret02}] seems to be
the most widely used concept of model selection in the past decade.
Nevertheless, DIC in the mixture context
lacks the theoretical foundations and its use remains controversial
[see comments and rejoinder on \citet{CeleuxForbesRobertTitterington06}].
For these reasons, \citet{Plummer08} suggested basing criteria for
model choice on penalized loss functions and cross-validating arguments.
For mixture models, in particular, he suggested using the penalized
expected deviance (PED). Since then, it has been successfully exploited
in several applications [\citet{KomarekNMix,CabralLachosMadruga11,delaFeRodriguetet11}], and we shall
use it
as well to choose the number of clusters.

The penalized expected deviance is defined as
%
\begin{equation}
\label{eqPED} \PED= \EV\bigl\{D(\psib, \thetab) | \yb\bigr\} +
p_{\mathrm{opt}},
\end{equation}
where $D(\psib, \thetab) = -2 \log\{L(\psib, \thetab) \}$ is the
observed data deviance of the model,
and its posterior mean $\EV\{D(\psib, \thetab) | \yb\}$ (expected
deviance) is easily estimated from
the MCMC sample. Further, the $p_{\mathrm{opt}}$ part of equation (\ref
{eqPED}) is the penalty term called optimism,
which can be estimated by using the two parallel MCMC chains and
importance sampling [\citet{Plummer08}].

%
\begin{table}[b]
\tablewidth=200pt
\caption{{\PBCsub} data. Penalized expected deviance for models with
$K=1, 2, 3, 4$ clusters}
\label{tabPBC06}
\begin{tabular*}{200pt}{@{\extracolsep{\fill}}lccc@{}}
\hline
$\bolds{K}$ & \textbf{\textsf{PED}} & $\bolds{\widehat{\EV}\{D(\psib,
\thetab) | \yb\}}$ & $\bolds{\widehat{p}}_{\mathbf{opt}}$ \\
\hline
1 & 14,277.9 & 14,241.8 & \phantom{00}36.1 \\
2 & \textbf{14,164.1} & 14,088.3 & \phantom{00}75.8 \\
3 & 14,183.1 & 14,057.1 & \phantom{0}126.0 \\
4 & 22,405.1 & 17,244.4 & 5160.8 \\
\hline
\end{tabular*}
\end{table}

\subsection{Number of clusters in the PBC910 data}\label{subsecselectKPBC}
Table~\ref{tabPBC06} shows calculated values of the penalized expected
deviance for the {\PBCsub} data and
models with $K=1,2,3,4$ clusters. It shows that the two-component model
fits the data clearly better than
a model with just a single cluster. On the other hand, the three
clusters are already too much for these data.
Even though the expected deviance of the three-component model is lower
than that of the two-component model,
the decrease of the expected deviance is overcome by the penalization
for the additional component.

The conclusion that the third cluster is redundant for our application
is also supported by the fact that when
a three-component model is fitted to the {\PBCsub} data, the two
components almost coincide with the mixture components
from the $K=2$ model and the estimated weight of the additional
component is only $\widehat{w}_3 = 0.021$,
and only three patients are allocated here using the rule based on
a maximal value of the posterior component probability.


\section{A simulation study}\label{secSimul}

\subsection{Simulation setup}
The setup of the simulation study was motivated by the {\PBCsub} application
and the data were generated according to the model (\ref{eqmodelPBC}).
For each subject $i$ and each marker $r$, $n_{i,r}=4$ visit times were
generated,
with the first visit time being equal to 0 and the remaining three
visit times being
generated from uniform distributions on intervals $(170, 200)$, $(350,
390)$, $(710, 770)$ days, respectively.
The covariate $t_{i,r,j}$ in (\ref{eqmodelPBC}) was the visit time in months.
The GLMM related parameters were equal: $\sigma_1 = \sqrt{\phi_1} =
0.3$, $\alpha_3 = 0.05$,
and we tried two values for the true number of clusters: $K=2$ and
$K=3$. In the two-cluster data,
both mixture weights were rather high: $\wb= (0.6, 0.4\tra{)}$,
whereas in the three-cluster data, a small third component with $w_3 = 0.06$
was created by splitting the second component of the two-cluster
setting, leading to the weights $\wb= (0.60, 0.34, 0.06\tra{)}$.
To make the differences between the clusters less obvious, not all
elements of the mixture means varied across the clusters;
see Table~\ref{tabsim}. Namely, in data with $K=2$, both clusters
shared the same value of the mean slope of the Gaussian response
and also the same value of the mean intercept of the Poisson response.
There were also some differences introduced
across the mixture covariance matrices; see Table~D.1 
in the Supplement [\citet{glmmClustWeb}].
Example data sets generated according to considered simulation settings are
also shown on Figures~D.1 and D.2 
of the Supplement.\looseness=-1

%
\begin{sidewaystable}
\tablewidth=\textheight
\tablewidth=\textwidth
\caption{Simulation study: \textup{(a)} proportions of models selected
with $K=1, 2, 3, 4$ using the PED criterion;
\textup{(b)} total classification error rate from a model with
correctly specified $K$; \textup{(c)} true values of mixture weights
and mixture means;
\textup{(d)} square roots of the mean squared errors (MSE), where the
calculated MSE is based on posterior means
as parameter estimates. For each parameter, the reported MSE is the
average MSE over the $K$ mixture components.
The $N$ gives the true number of subjects in each cluster}
\label{tabsim}
{\fontsize{7.6}{9.6}{\selectfont
\begin{tabular*}{\textwidth}{@{\extracolsep{\fill
}}lccd{2.4}cd{2.4}d{2.2}cd{2.1}d{2.0}ccc@{}}
\hline
& \multicolumn{7}{c}{} & & \multicolumn{4}{c}{\multirow
{2}{80pt}{\centering\textbf{Proportion (\%) of models selected with $K$}}}
\\[3pt]
\multicolumn{9}{c}{}& \multicolumn{4}{c@{}}{\hrulefill}\\
\textbf{Setting} & $\bolds{\wb}$ & \multicolumn{1}{c}{$\bolds{\mu
_{*,1}}$} & \multicolumn{1}{c}{$\bolds{\mu_{*,2}}$} &
\multicolumn{1}{c}{$\bolds{\mu_{*,3}}$} & \multicolumn{1}{c}{$\bolds{\mu
_{*,4}}$} & \multicolumn{1}{c}{$\bolds{\mu_{*,5}}$}
& \multicolumn{1}{c}{$\bolds{\alpha_3}$} & \multicolumn{1}{c}{\multirow
{2}{50pt}[10pt]{\centering\textbf{Classif. error rate (\%)}}}
& \textbf{1} & \textbf{2} & \textbf{3} & \textbf{4} \\
\hline
\multicolumn{1}{@{}l}{$K = 2$} & \multicolumn{7}{c}{\textit{True
values}} & &&&& \\
& 0.600 & 0.000 & 0.0100 & 5.00 & -0.0050 & -3.00 & 0.050 & &&&& \\
& 0.400 & 1.000 & 0.0100 & 5.00 & -0.0200 & -1.00 & & &&&&
\\[3pt]
& \multicolumn{7}{c}{\textit{Square root of the MSE}} & &&&& \\
\multicolumn{13}{@{}l}{Normal}\\
$N = (30, 20)$ & 0.169 & 0.237 & 0.0048 & 0.12 & 0.0039 & 1.84 & 0.029
& 15.8 & 67 & \textbf{33} & \phantom{0}0 & 0 \\
\quad$(60, 40)$ & 0.064 & 0.150 & 0.0034 & 0.09 & 0.0018 & 0.86 & 0.018
& 7.8 & 15 & \textbf{85} & \phantom{0}0 & 0 \\
\quad$(120, 80)$ & 0.037 & 0.066 & 0.0022 & 0.05 & 0.0010 & 0.46 &
0.014 & 5.8 & 1 & \textbf{99} & \phantom{0}0 & 0
\\[3pt]
\multicolumn{13}{@{}l}{MVT$_5$}\\
$N = (30, 20)$ & 0.209 & 0.564 & 0.0134 & 0.36 & 0.0051 & 3.33 & 0.029
& 20.6 & 64 & \textbf{35} & \phantom{0}1 & 0 \\
\quad$(60, 40)$ & 0.128 & 0.433 & 0.0103 & 0.47 & 0.0035 & 2.17 & 0.020
& 10.8 & 17 & \textbf{76} & \phantom{0}7 & 0 \\
\quad$(120, 80)$ & 0.102 & 0.378 & 0.0049 & 0.16 & 0.0020 & 1.47 &
0.013 & 8.7 & 0 & \textbf{82} & 18 & 0
\\[6pt]
\multicolumn{1}{@{}l}{$K = 3$} & \multicolumn{7}{c}{\textit{True
values}} & &&&& \\
& 0.600 & 0.000 & 0.0100 & 5.00 & -0.0050 & -3.00 & 0.050 & &&&& \\
& 0.340 & 1.000 & 0.0100 & 5.00 & -0.0200 & -1.00 & & &&&& \\
& 0.060 & 1.300 & -0.0300 & 5.50 & 0.0000 & -2.00 & & &&&&
\\[3pt]
& \multicolumn{7}{c}{\textit{Square root of the MSE}} & &&&& \\
\multicolumn{13}{@{}l}{Normal}\\
$N = (30, 17, 3)$ & 0.154 & 0.444 & 0.0204 & 0.33 & 0.0073 & 5.04 &
0.027 & 26.5 & 80 & 18 & \phantom{0}\textbf{2} & 0 \\
\quad$(60, 34, 6)$ & 0.088 & 0.360 & 0.0165 & 0.26 & 0.0054 & 3.07 &
0.018 & 17.4 & 36 & 50 & \textbf{14} & 0 \\
\quad$(120, 68, 12)$ & 0.048 & 0.260 & 0.0126 & 0.19 & 0.0034 & 1.58 &
0.015 & 10.1 & 2 & 59 & \textbf{38} & 1
\\[3pt]
\multicolumn{13}{l}{MVT$_5$}\\
$N = (30, 17, 3)$ & 0.113 & 0.563 & 0.0237 & 0.52 & 0.0078 & 3.99 &
0.021 & 23.8 & 66 & 31 & \phantom{0}\textbf{3} & 0 \\
\quad$(60, 34, 6)$ & 0.122 & 0.424 & 0.0211 & 0.31 & 0.0058 & 2.66 &
0.021 & 18.8 & 24 & 59 & \textbf{17} & 0 \\
\quad$(120, 68, 12)$ & 0.056 & 0.353 & 0.0196 & 0.22 & 0.0048 & 1.56 &
0.013 & 9.7 & 2 & 44 & \textbf{50} & 4 \\
\hline
\end{tabular*}
}}
\end{sidewaystable}

To examine the performance of our method in situations when there is
misspecification in the random effects distribution,
we simulated data not only under the normal distribution of random
effects, but also under the
shifted-scaled multivariate $t$-distribution with five degrees of freedom
(MVT$_5$).
For each setting ($K$, distribution of random effects), we tried three
values of sample sizes with total numbers of subjects
being $50, 100, 200$. For each setting and each sample size, 100 data
sets were generated.
The posterior inference for each data set was based on 10,000
iterations of 1:100 thinned MCMC obtained after a burn-in period
of 1000 iterations.

\subsection{Parameter estimates}
The left block of Table~\ref{tabsim} shows the square roots of the mean
squared errors (MSE) in the estimation
of the most important model parameters (mixture weights and means, GLMM
fixed effects) provided that
the number of clusters, $K$, is correctly specified. As parameter
estimates, we considered the posterior means,
and for a particular parameter the reported MSE is the average MSE over
the parameter values from all $K$ components.
Detailed results also providing the bias and the standard deviations of
the posterior means are given
in Tables D.3---D.6 
of the Supplement [\citet{glmmClustWeb}].

With normally distributed random effects, the posterior means of model
parameters seem to provide consistent estimates
of the model parameters. The same can also be concluded\vadjust{\goodbreak} when the true
distribution of random effects is MVT$_5$. Nevertheless,
not surprisingly, the convergence of the parameter estimates to their
true values is in general slower in this case.

\subsection{Classification error rates}
With respect to classification, one of the most important measures is
the classification error rate reported in the ninth
column of Table~\ref{tabsim}. More detailed results, also showing
conditional (given the cluster) classification error rates,
are shown in Table D.2 
of the Supplement [\citet{glmmClustWeb}]. Among other things, we point
out that even with
incorrectly specified distribution of random effects, the
classification error rates do not differ considerably from the
error rates obtained with data for which the random effects
distribution was correctly specified, and even with a moderate sample size
of 200 subjects, the achieved error rate is as low as 10\%.

\subsection{Selection of a number of clusters}
Performance of the penalized expected deviance as a tool for the
selection of a number of clusters is illustrated by the right block
of Table~\ref{tabsim}. For each simulated data set, we estimated four
models, each of them under the assumption of a different number of clusters,
namely, $K=1, 2, 3, 4$, and calculated the corresponding PED values.
Table~\ref{tabsim} shows proportions of data sets for which
a particular number of clusters was selected by minimizing the PED. For
each simulation setting, bold numbers indicate
proportions of data sets with a correctly selected number of clusters.

For data sets composed of two clusters both having rather high weights,
the probability of a selection of a correct model increases with the
sample size,
practically reaching 100\% with $N = 200$, and normally distributed
random effects. The situation is slightly worse when the true
distribution of random effects is MVT$_5$. Nevertheless, the results
are also rather satisfactory in this case.

When there were three clusters present, with one of the clusters having
a rather small weight of 0.06, the probability
of a correct selection of a number of clusters also increases with the
sample sizes (for both normally and MVT$_5$ distributed
random effects). However, it is much slower than in the case of two
clusters of almost equal size.
Nevertheless, we point out that already for the moderate sample size
of 200 subjects, in 97\% and 94\% of cases, respectively, the number of
clusters indicated by the value of PED differs by at most one from the correct
value of three.

\section{Discussion}\label{secDiscuss}
In this paper we have proposed a method for classification of subjects
on the basis of longitudinal measurements
of several outcomes of a different nature which, according to the best
of our knowledge, has not been considered in the literature yet.
The clustering procedure relies on a classical GLMM specified for each
marker, whereas
possible dependence across the values of different markers is captured
by specifying a joint distribution of all random effects.
In contrast to a classical assumption,
we assume a normal mixture in the random effects distribution which is
the core classification component of the proposed model.
Although other choices of the basis distributions
could be considered as well, we would like to stress the fact that with
the Bayesian approach used here, the normality assumption
only corresponds to a particular choice of just one component of the
overall prior distribution which is updated by the data. For example,
the posterior predictive distribution for random effects is given by
\begin{displaymath}
p_{\mathrm{pred}}(\bb) = \int\Biggl\{\sum_{k=1}^K
w_k \varphi(\bb; \mub_k, \Db_k) \Biggr\} p
(\thetab| \yb) \,d\thetab,
\end{displaymath}
which in general is no longer a Gaussian mixture. In a similar way, the
posterior distribution enters the calculation of
the posterior component probabilities (\ref{eqpiik}) while taking into
account the uncertainty in the specification
of the distribution of random effects. Last but not the least, the
simulation study also suggests that,
at least in situations when the true random effects
distribution is symmetric with heavier tails, assuming a priori
a normal random effects distribution does not have
any crude impact on classification error rates.

Being within the Bayesian framework, it is also relatively easy to
calculate not only point estimates of the individual component
probabilities, but also corresponding credible intervals which can be
subsequently used to evaluate uncertainty in the
pertinence of a particular subject in a specific cluster. Such
uncertainty is only rarely evaluated in similar situations.
Finally, we adapted recently published methodology for model comparison
based on the concept of a penalized expected deviance
to explore the optimal number of clusters.\looseness=-1

Further, we point out that even though we classified in this paper only
those subjects who were also used to draw the posterior inference
on the model parameters, our procedure can also be used to classify
a new subject with a value of observed longitudinal markers equal to
$\yb_{\mathrm{new}}$
and unknown allocation $u_{\mathrm{new}}$.
Indeed, given the posterior sample $\Samp_M$ from $p (\psib, \thetab |
\yb)$, the component probabilities
$p_{\mathrm{new},k} (\psib^{(m)}, \thetab^{(m)} )$ and estimated posterior
component probabilities $\widehat{\pi}_{\mathrm{new},k}$ $(k=1,\ldots,K)$
can be calculated using expressions (\ref{eqpik}) and (\ref{eqpiik}),
and then used for classification. Finally, it is worth mentioning
that discriminant analysis, where a training data set with known
cluster allocation is available, would also be possible
with only slightly modified methodology where separate posterior
samples would be drawn using the data from the various clusters
and then used to calculate the component probabilities.

For practical analysis, we extended the R package \texttt{mixAK} [\citet{KomarekNMix}]
to cover the methodology proposed in this paper. The package is freely
available from the \textit{Comprehensive R Archive Network}
at \href{http://cran.r-project.org/}{http://cran.r-project.org/}.


\section*{Acknowledgments}
We thank Mr. Steven Del Riley for his help with English corrections of
the manuscript.
Last but not least, we thank two anonymous referees, an Associate
Editor and the Editor for
very detailed and stimulative comments to the earlier versions of this
manuscript that led
to its considerable improvement.

\begin{supplement}[id=suppA]
\stitle{Appendices}
\slink[doi]{10.1214/12-AOAS580SUPP}  
\sdatatype{.pdf}
\sfilename{aoas580\_supp.pdf}
\sdescription{The pdf file contains (A)~more detailed description of
the assumed prior distribution for model parameters,
giving also some guidelines for the selection of the hyperparameters
to achieve a weakly informative prior distribution;
(B) more details on the posterior distribution and the sampling MCMC algorithm;
(C) additional information to the analysis of the Mayo Clinic PBC data;
(D) more detailed results of the simulation study.}
\end{supplement}

%

\printaddresses

\end{document}